\documentclass[12pt,epsf,epsfig,wrapfig]{article}
\usepackage{graphicx}
\begin{document}

\begin{center}
{\bfseries TRANSVERSITY PHYSICS AT COMPASS}
\vskip 5mm
F. Bradamante~\footnote{ \noindent
Invited talk at the 11th International Workshop on High Energy 
Spin Physics (DUBNA-SPIN-05), Dubna, Russia, September 27 - October 1, 2005}\\
\vskip 3mm
{\small
{\it
University of Trieste and INFN Trieste\\
E-mail: Franco.Bradamante@trieste.infn.it
}}
\vskip 5mm
on behalf of the COMPASS Collaboration
\end{center}

\vskip 5mm

\begin{abstract}
Transverse spin physics is an important part of the scientific
programme of the COMPASS experiment at CERN, which started taking
data in 2002, scattering 160 GeV/c muon beam on a polarized
$^6$LiD target.
The analysis of the data taken with the target polarized 
orthogonally to the muon beam direction has allowed to measure
for the first time the Collins and Sivers asymmetries
of the deuteron.
Both for the positive and the negative hadrons produced
in semi-inclusive DIS the measured asymmetries are small
and, within errors, compatible with zero: results on part of
the accumulated statistics have already been published.
Two-hadron asymmetries and $\Lambda$ polarization transfered
from the struck quark are also being investigated, and preliminary
results on the data collected in the years 2002 and 2003
are given. 
\end{abstract}

\vskip 8mm 
%
\section{Introduction}
The importance of transverse spin effects at high energy
in hadronic physics has grown up steadily since the
discovery in 1976 that
$\Lambda$ hyperons produced in $pN$ interactions exhibit
an anomalously large transverse polarization~\cite{Bunc76}. 
Nowadays transversity is the subject of intense theoretical 
activity~\cite{BDR02},
an important part of the scientific programme of the HERMES experiment
at DESY, of the COMPASS experiment at CERN, and of the RHIC experiments
at BNL, and the topical content of important international
workshops~\cite{como}.

To completely
specify the quark structure of the nucleon at the twist-two level, 
the transverse spin distributions $\Delta_T q(x)$ must be added to the momentum
distribution $q(x)$ and to the helicity distribution 
$\Delta q(x)$~\cite{JaJi91}.
If the quarks are collinear with the parent nucleon (no intrinsic $k_T$),
or after integration over $k_T$,
these three distributions exhaust the information
on the internal dynamics of the nucleon.

The transversity distributions $\Delta_T q(x)$ have never been measured, since 
they are chirally-odd and therefore absent in
inclusive DIS, the usual source of information on the nucleon partonic 
structure. 
As suggested in Ref.~\cite{Coll93}, they  may instead be
extracted from measurements of the spin asymmetries in cross-sections 
for semi-inclusive DIS (SIDIS)
between leptons and transversely polarized nucleons, in which 
a hadron is also detected in the final state. In
such processes the measurable asymmetry $A_{Coll}$ (``Collins asymmetry'')
is due to the combined effect 
of $\Delta_T q(x)$ and another chirally-odd function, $\Delta^0_T D_q^h$, which
describes the spin dependent part of the
hadronization of a transversely polarized quark $q$ in
a hadron $h$.

A different mechanism has also been suggested in the past~\cite{Siv90} as a 
possible cause of
a spin asymmetry  in the cross-section of SIDIS 
between leptons and transversely polarized nucleons.
Allowing for an intrinsic $k_T$ dependence of the quark distribution
in a nucleon, a left-right asymmetry could be induced in such a distribution
by a transverse nucleon polarization, thus causing an asymmetry 
$A_{Siv}$ (the ``Sivers
asymmetry'') in the quark fragmentation hadron with respect to the nucleon 
polarization.
These two mechanisms can be measured
separately in SIDIS on a transversely polarized nucleon~\cite{colsiv}.

According to Collins, 
the fragmentation function of a quark of flavour $q$ into a hadron $h$
can be written as 
\begin{eqnarray}\label{eq:Dt}
D_q^h(z, \vec{p}_T) = D_q^h(z, p_T) + \Delta_T^0 D_q^h(z, p_T) 
\cdot \sin\Phi_C
\end{eqnarray}
where
$\vec{p}_T$ is the final hadron transverse momentum
with respect to the quark direction (e.g. the virtual photon direction),
$z = E_h / (E_{l}-E_{l'})$ is the fraction of available energy 
carried by the hadron ($E_h$ is the hadron energy, $E_{l}$ is 
the incoming lepton
energy and $E_{l'}$ is the scattered lepton energy).
The ``Collins angle'' $\Phi_C$ is conveniently defined in a system in
which the z--axis is the  virtual photon direction
and the x--z plane is the muon
scattering plane (see Fig.~1). In this reference system
$\Phi_C= \phi_h -\phi_{s'}$, where
$\phi_h$ is the hadron azimuthal angle, and $\phi_{s'}$ is the azimuthal angle
of the transverse spin of the struck quark, as illustrated in 
Fig.~1. Since $\phi_{s'}=\pi-\phi_s$, with
$\phi_s$ the transverse spin of the initial quark (nucleon), it is also
$\Phi_C = \phi_h +\phi_s - \pi$, i.e. $\sin \Phi_C = - \sin(\phi_h +\phi_s)$.
\begin{figure}[bt] %
\begin{center}
\includegraphics[width=10.cm]{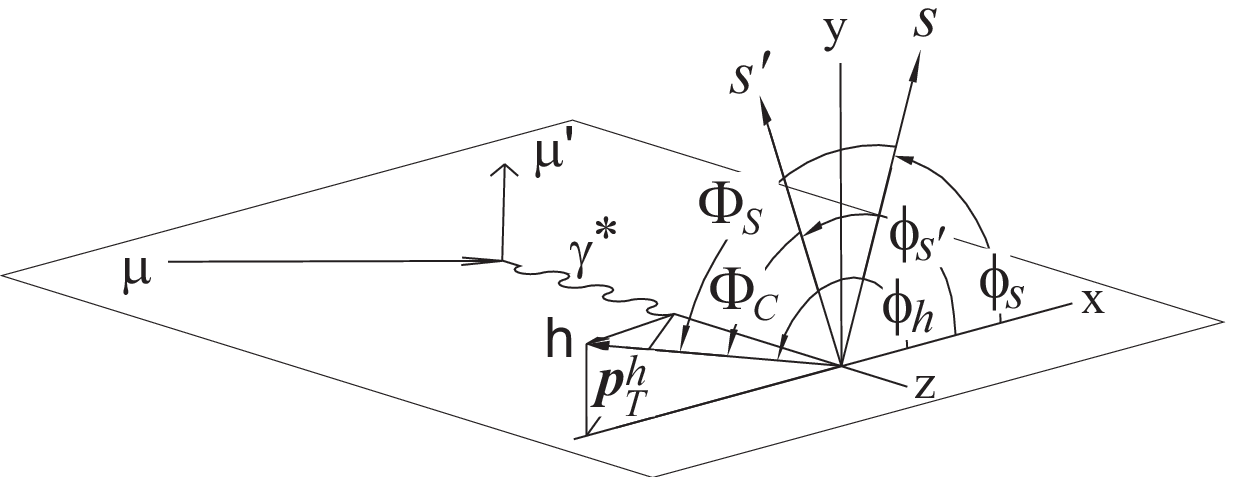}\\
\vspace*{.4cm}
{\small{\bf Figure 1.} Definition of the Collins and Sivers angles.}
\end{center}
\end{figure}
At leading order 
in the collinear case $A_{Coll}$ can be written as 
\begin{eqnarray}
A_{Coll} = \frac {\sum_q e_q^2 \cdot \Delta_T q \cdot \Delta_T^0 D_q^h}
{\sum_q e_q^2 \cdot q \cdot D_q^h}
\label{eq:collass}
\end{eqnarray}
where $e_q$ is the quark charge.
The quantity $\Delta_T^0 D_q^h$ can be obtained by
investigating the fragmentation of a polarized quark
$q$ into a hadron $h$.

In the approach of Sivers, 
allowing for an intrinsic $\vec{k}_T$ dependence of the quark distribution
in a nucleon, a left-right asymmetry could 
be induced in such a distribution by a transverse nucleon polarization, 
$q_{T}(x,\vec{k}_T)= q(x,|\vec{k}_T|^2) + \Delta_0^T q(x,|\vec{k}_T|^2)
\cdot \sin \, \Phi_{ S}$, where $\Phi_{ S}= \phi_h -\phi_s \neq \Phi_{C}$ 
is the ``Sivers angle''.
Neglecting the hadron transverse momentum with respect to the
fragmenting quark, this $\vec{k}_T$ dependence could cause the ``Sivers
asymmetry'' 
\begin{eqnarray}
A_{Siv} & = & \frac {\sum_q e_q^2 \cdot \Delta_0^T q \cdot D^h_q}
{\sum_q e_q^2 \cdot q \cdot D_q^h} 
\label{eq:sivass}
\end{eqnarray}
in the distribution of the hadrons resulting from the quark
fragmentation with respect to the nucleon polarization which
could be revealed as a $\sin \Phi_{ S}$ modulation
in the distribution of produced hadrons.

\section{Single hadron asymmetries}
The COMPASS experiment is described in these Proceedings~\cite{rw} and in more
detail in Ref.~\cite{FB03}. 
Up to now, data have been taken on polarized deuterons, using a
polarized $^6$LiD target.
The polarized target consists of two  cells, each 60 cm long, 
located along the beam line, one after the other, in two separate RF cavities. 
Data are taken simultaneously on the two target cells, which are oppositely 
polarized. 
The target magnet can provide both a solenoid field (2.5~T),
for longitudinal polarization measurements, and a dipole field
(0.5~T) used for adiabatic spin rotation and for the transversity measurements.
Correspondingly, the target polarization can be oriented either longitudinally
or transversely to the beam direction.
Polarizations of 50\% have been reached routinely with the $^6$LiD target,
which has a favourable dilution factor $f \simeq 0.4$, since $^6$LiD
basically consists of a deuteron plus an $^4$He core.
When operating in the transverse polarization mode the polarization is 
reversed after 4-5 days by changing the 
RF frequencies in the two target cells.
Typical polarization values for the transverse running are 50\% in
the first period of data taking, and 40\% in the subsequent periods,
after polarization reversal.

About 20\% of the total beam-time in 2002, 2003 
and 2004 was devoted to the run 
with the transversely polarized deuteron target. The accumulated sample of 
2002 data with transverse polarization of the target comprises 
$6 \cdot 10^9$ events. 
The amount of data collected in 2003 is  a factor of two larger than 
in 2002, and that collected in 2004 is  a factor of two larger 
than in 2003.
The single hadron analysis of the 2003 and 2004 data is not yet finalized,
and I will report only on the 2002 results.

In 2002 the data were taken during two separate periods, and
in each period data were taken with  two different
orientations of the target cells. 

To select semi-inclusive events, an incoming and scattered muon  
(primary vertex) plus at least one charged 
hadron from this vertex were required. 
Muon identification was performed with muon filters, consisting of a large 
amount of material to be traversed and a suitable tracking system. 
To select DIS events, the kinematic cuts 
$Q^2>1$ (GeV/c)$^2$, $W > 5$ GeV/c$^2$ and $0.1< y <0.9$ were 
applied to the data. 
The upper limit on $y$ was applied to keep radiative corrections small. 
In addition the transverse momentum cut $p_T > 0.1$ GeV/c was applied to 
unambiguously calculate angles. Asymmetries have been extracted for the 
charged hadron with the highest energy (leading, $z > 0.25$) 
as well as for all 
detected charged hadrons with $z > 0.2$.
The identification provided by the RICH was not used in this analysis.

Collins and Sivers asymmetries were fitted separately. 
The number of events for each cell and for each orientation of the
target polarization can be written as
\begin{eqnarray}
N_h^{\uparrow (\downarrow)}(\Phi_{C/S}) & \propto & 
(1 \pm \epsilon_{C/S} \cdot \sin \Phi_{C/S}), 
\end{eqnarray}
where $\epsilon_{C/S}$  is the amplitude of the experimental asymmetry, 
the arrows refer to the two orientations of the target polarization,
and $\Phi_{C/S}$ is evaluated as if the spin of the target always
pointed upward, along the vertical direction.

The quantities $\epsilon_{C}$ and $\epsilon_{S}$ 
can be written as a function of the 
Collins and Sivers asymmetries:
\begin{eqnarray}
\epsilon_{C} & = & A_{Coll} \cdot P_T \cdot f \cdot D_{NN}\, , \nonumber \\
\epsilon_{S} & = & A_{Siv} \cdot P_T \cdot f  \, ,
\label{eq:colsiv}
\end{eqnarray}
where $P_T$ is the target polarization, 
$D_{NN} = (1 - y)/(1 - y + y^2/2)$ 
is the transverse spin transfer coefficient from the initial to the 
struck quark, and 
$f$ is the target dilution factor. 
The asymmetry $\epsilon_{C/S}$ is fitted separately for the two target 
cells with two 
opposite spin orientations from the expression:
\begin{eqnarray}
\epsilon_{C/S} \cdot \sin \Phi_{C/S} & = &
\frac{N_h^{\uparrow}(\Phi_{C/S}) - r \cdot N_h^{\downarrow}(\Phi_{C/S})}
{N_h^{\uparrow}(\Phi_{C/S}) + r \cdot N_h^{\downarrow}(\Phi_{C/S})}
\end{eqnarray}
where $r = N_{h,tot}^{\uparrow} / N_{h,tot}^{\downarrow}$ 
is a normalization factor and corresponds to the 
ratio of the total number of events with two target polarization orientations. 
The  results from the two target cells and for the 
two data-taking periods have been averaged, after checking
their statistical compatibility.

\begin{figure}[bt] %
\begin{center}
\includegraphics[width=13.cm]{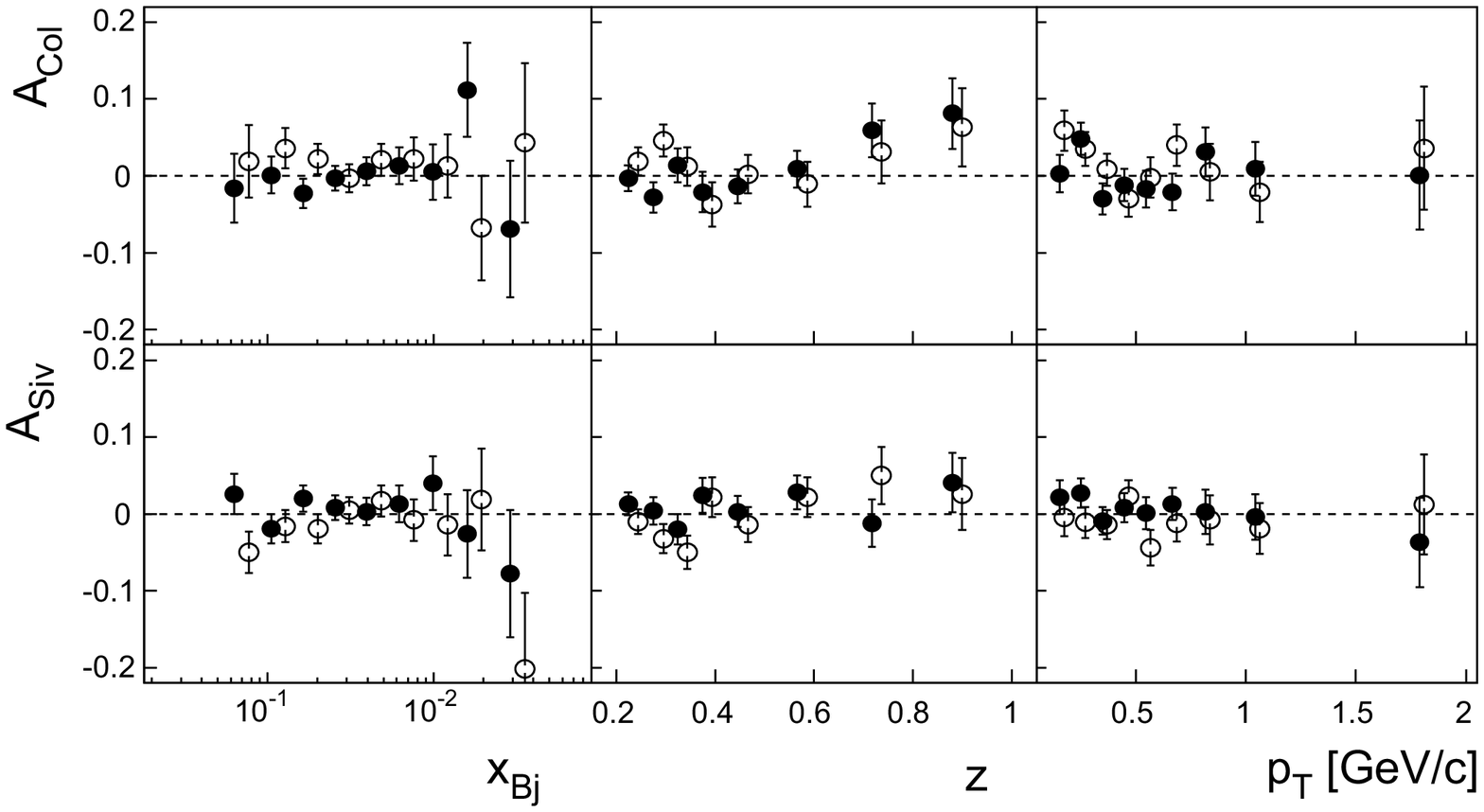}\\
\end{center}
\vspace*{-0.4cm}
{\small{\bf Figure 2.} Collins and Sivers asymmetry for positive (full points) 
and negative
(open points) hadrons as a function of $x$, $z$ and $p_T$.}
\end{figure}
The resulting asymmetries are plotted against the kinematic variables 
$x$, $z$ and $p_T$ in Fig.~2 for all hadrons.
These results, which refer to the 2002 run, have already been 
published~\cite{ourPRL}.
Similar results have been obtained in the case of leading 
hadrons. 
Only statistical errors are shown in Fig.~2:
systematic errors have been shown to be 
smaller  than statistical ones.
Full points correspond to the positively charged hadrons and 
open points correspond to the negatively charged hadrons.
As apparent from Fig.~2 
both the Collins and Sivers asymmetries are 
small and compatible with zero. 
This might either hint to a cancellation between proton and neutron or to a
too small Collins mechanism. However, if the Collins function 
$\Delta_T^0 D^h_q (z, p_T )$ is not zero and as large as 
indicated from the preliminary 
results by the BELLE Collaboration~\cite{belle}, then our data
provide evidence for 
cancellation in the isoscalar target.

The HERMES transversity data on protons and our transversity data on deuterons
already allow for a combined analysis aiming at the extraction
of the Sivers functions and of the transversity distributions.
Within the limited statistical accuracy of the published data,
a few global analysis have already been performed, and the observed 
phenomena can be described in a unified 
scheme~\cite{Anselmino:2005ea,Vogelsang:2005cs}.

\section{Two-hadron asymmetries}

Another process has been suggested to address transversity, namely
semi-inclusive DIS where at least two-hadrons are 
observed in the final state. 
Here the cross-section at leading twist 
can be parametrized in terms of the convolution of transversity with 
an interference fragmentation function 
$H_1^{< \hspace{-.8ex} {\scriptscriptstyle )}}(Z,M^2_{inv})$,
which is also chirally-odd.  
If a pair of hadrons is the result of the fragmentation of a transversely 
polarised quark, an 
asymmetry $A_{\phi_{RS}}$ depending on the angle between the scattering plane 
and the 2 hadron plane is expected: 
\begin{eqnarray}
A_{\phi_{RS}} & = &
\frac{\Sigma_q e^2_q \Delta_T q(x) \cdot
H_1^{< \hspace{-.8ex} {\scriptscriptstyle )}}(Z,M^2_{inv})}
{\Sigma_q e^2_q q(x) D^h_q(Z,M^2_{inv})}.
\label{asym_eq}
\end{eqnarray}
Here $Z$ is the 
sum of scaled hadron energies ($Z=z_{h1}+z_{h2}$) and $M_{inv}$ is 
the invariant mass of two-hadrons. The sum is over the quark flavours $q$. 
The expected properties of the interference fragmentation function 
and suggestions on how to access it experimentally can be found
in several publications~\cite{t1,t2,t4,t6}. 

The asymmetry (\ref{asym_eq}) is related to the experimentally measured 
counting rate asymmetry, which is defined in the following way:
\begin{eqnarray}
A_m(\phi_{RS})& = &\frac{N^{\uparrow}(\phi_{RS})-rN^{\downarrow}(\phi_{RS})}
{N^{\uparrow}(\phi_{RS})+rN^{\downarrow}(\phi_{RS})} \nonumber \\
& = & A^{\sin \phi_{RS}}_{UT} 
\sin \phi_{RS}=
D_{NN}P_TfA_{\phi_{RS}}\sin\phi_{RS},
\end{eqnarray}
where $N^{\uparrow(\downarrow)}$ is the number of events for the target with 
the up (down) 
polarization orientation in the laboratory system, 
$r=N^{\uparrow}_{tot} / N^{\downarrow}_{tot}$ is the ratio of 
the total number of events (i.e. integrated over the angle $\phi_{RS}$) for 
the two-polarization orientations,  and
$D_{NN}$, $P_T$ and  $f$ have the same meaning as in 
formula~(\ref{eq:colsiv}).
The angle $\phi_{RS}$ is defined as $\phi_{RS}=\phi_R - \phi_{S'}$, 
where $\phi_{R}$ is 
the azimuthal angle of $\vec{R}_T$,  
which is the vector component of the difference of the two-hadron 
momenta 
$\vec{R}_h=(\vec{P}_{1}-\vec{P}_{2})/2$ perpendicular 
to their sum $\vec{P}_h=( \vec{P}_{1}+\vec{P}_{2}$); 
$\phi_{S'}$ is still the azimuthal angle of 
the struck quark spin. 
\begin{figure}[tb] %
\begin{center}
\includegraphics[width=8.cm]{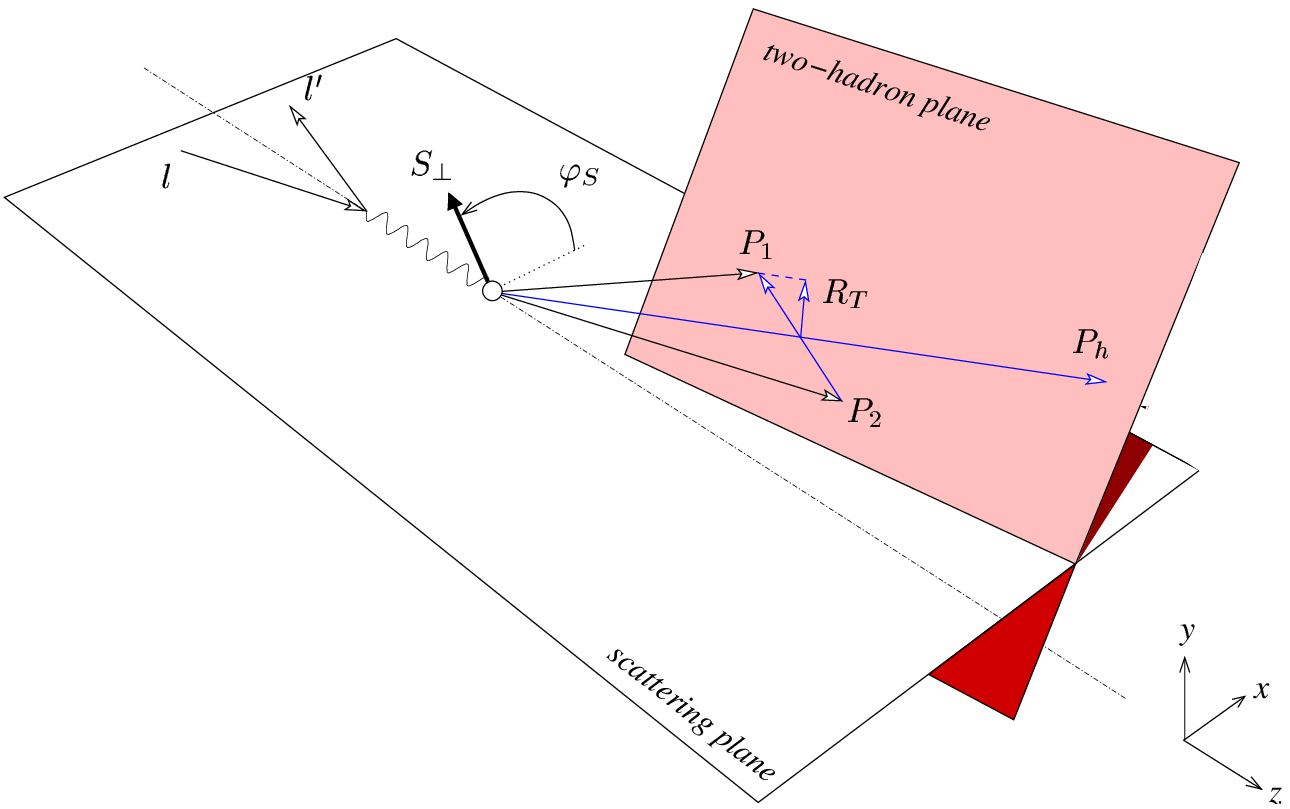}\\
\end{center}
{\small{\bf Figure 3.} Reference system and angles definitions for 
the two-hadron analysis~\cite{bacc}.
$\vec{S}_{\perp}$ is the vector component of the target quark spin
perpendicular to the virtual photon direction.}
\end{figure}
The reference system for the measurement is 
again defined by the scattering plane of the 
lepton  and the direction of the virtual photon,
 as shown in Fig.~3. 

The asymmetries $A^{\sin \phi_{RS}}_{UT}$ are obtained from the fit to the 
$A_m(\phi_{RS})$ distributions. 
Here we give preliminary results from the data collected during the years 
2002 and 2003.

The event selection is basically the same as for the single hadron analysis,
plus the requirement  that the events contain at least one reconstructed 
hadron pair with oppositely charged hadrons. 
In the calculation of $\vec{R}_T$, we always take as hadron 1
the positively charged hadron.
If more hadrons are present in an event,
$\phi_{RS}$ is evaluated for all pairs of positive and negative hadrons.
Current fragmentation region is guaranteed by cuts on $z_{h1,2}>0.1$ and $x_F>0.1$ 
for each hadron. 
In order to remove exclusive mesons a cut on $Z<0.9$ is applied. 
An additional cut on $R_T$ ($R_T> 50$ MeV/c) is performed in order to 
have well defined 
angles.

\begin{figure*}[h!]
\begin{center}
\vspace*{-.4cm}
\begin{tabular}{c}
\includegraphics[width=10.cm]{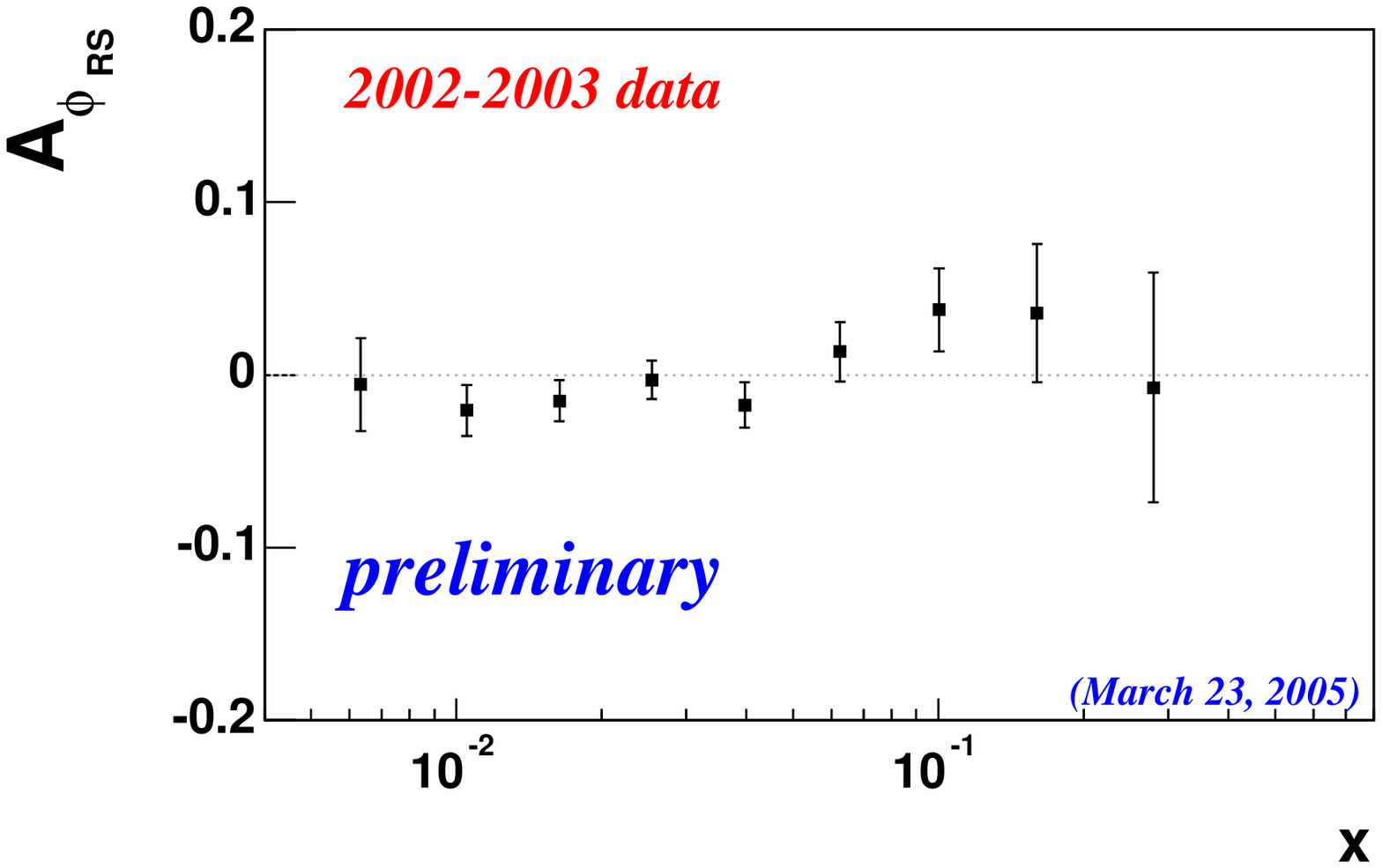}
\vspace*{-0.9cm}\\
\includegraphics[width=10.cm]{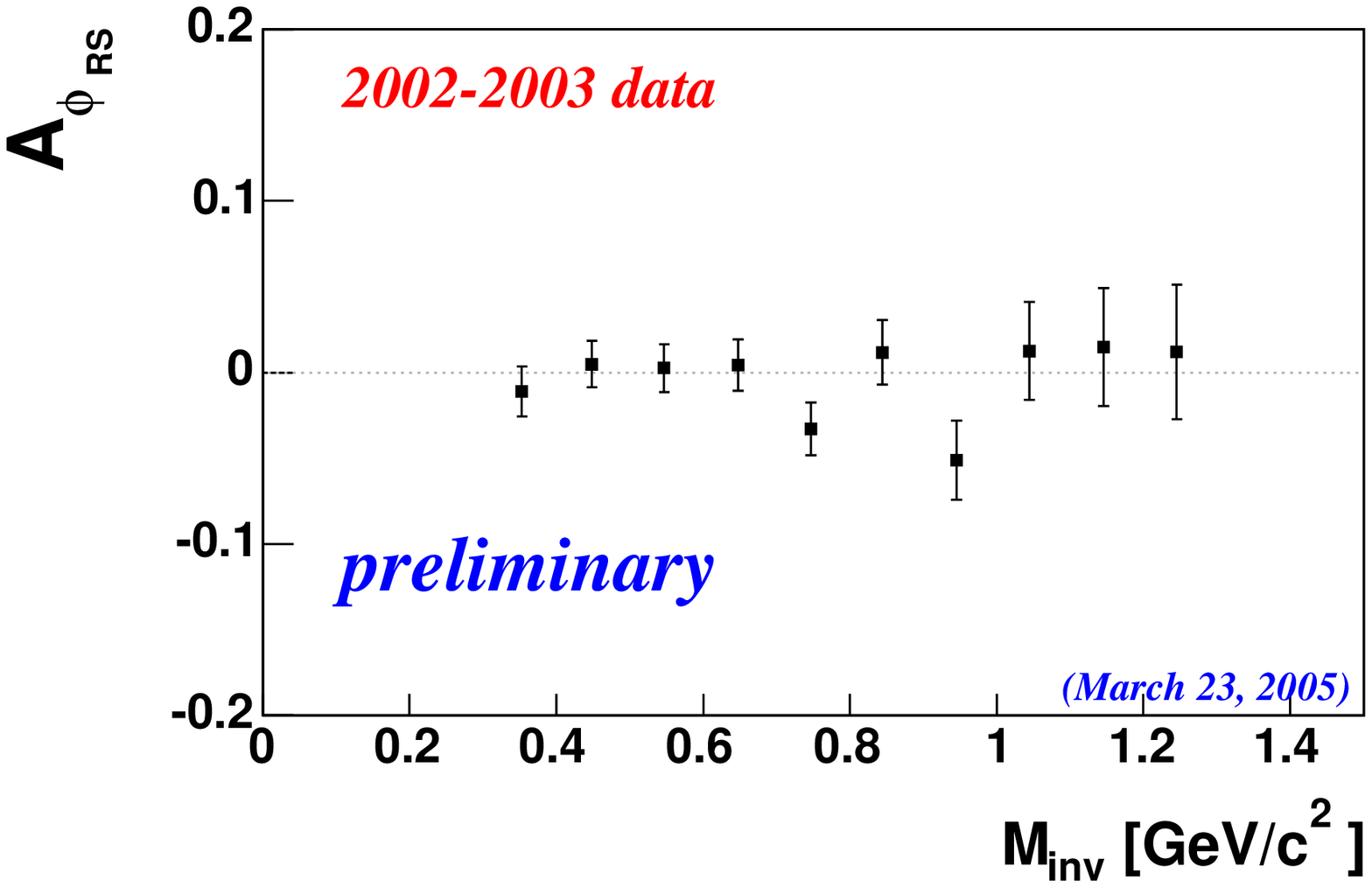}
\end{tabular}
\end{center}
\vspace*{-.7cm}
{\small{\bf Figure 4.} Two-hadron asymmetry $A_{\phi_{RS}}$
as a function of $x$ (top),
of the two-hadron invariant mass $M_{inv}$ (bottom).} 
\end{figure*}
The resulting asymmetries have been shown for the first time
at DIS05~\cite{rainer}, and are given in Fig.~4 as functions 
of the variables $x$ and $M_{inv}$.
For the invariant mass calculation, all 
hadrons are assumed to be pions. 
The results are compatible with zero. The indicated errors are statistical. 
The size of the 
systematics errors were estimated by evaluating ``false asymmetries'' of the 
data.
The extracted false asymmetries are compatible with zero and their statistical 
errors are 
of the same size as the statistical uncertainties of the physics result. 
\begin{figure}[h!]
\begin{center}
\includegraphics[width=9.5cm]{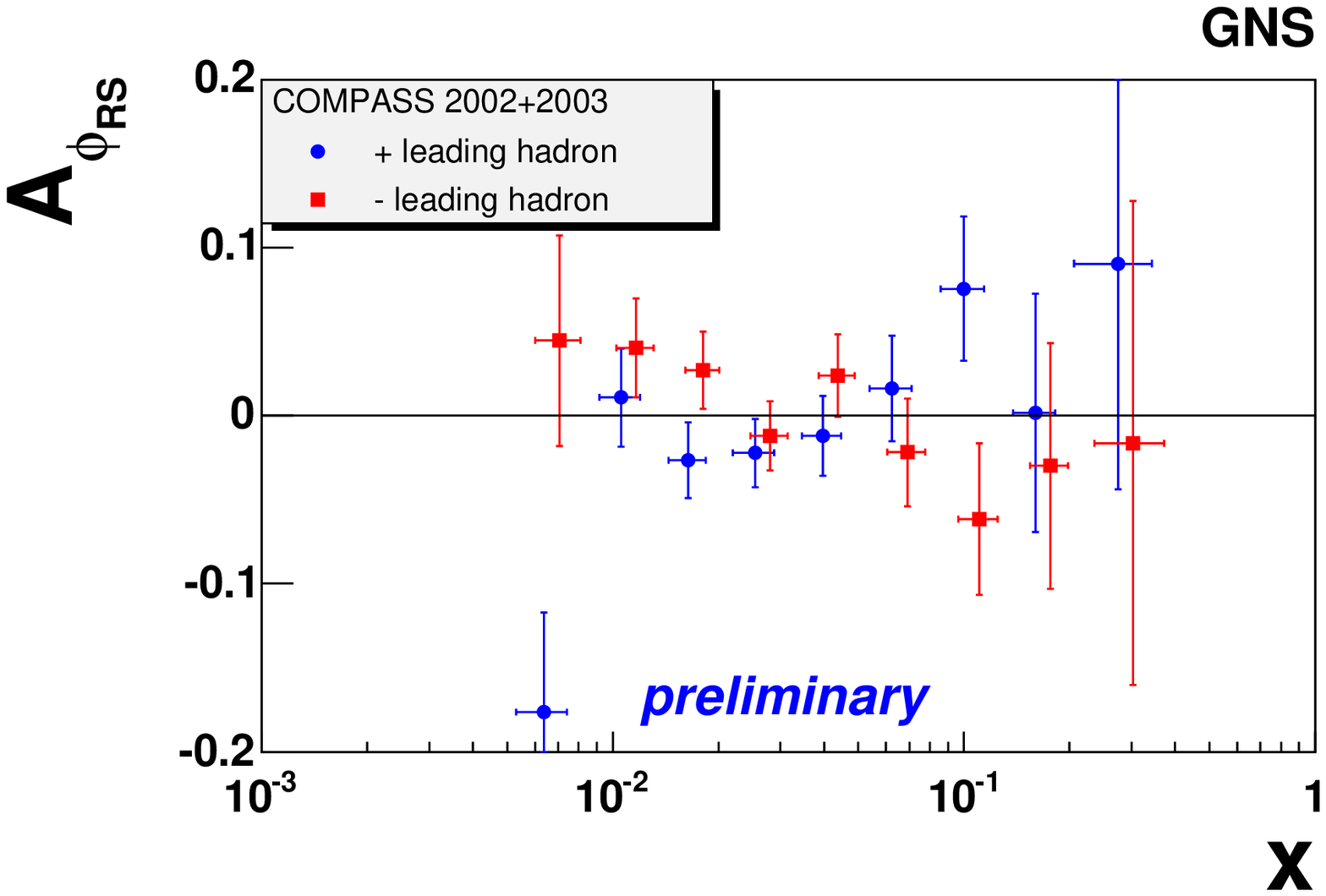}
\vspace*{-.4cm}\\
\includegraphics[width=9.5cm]{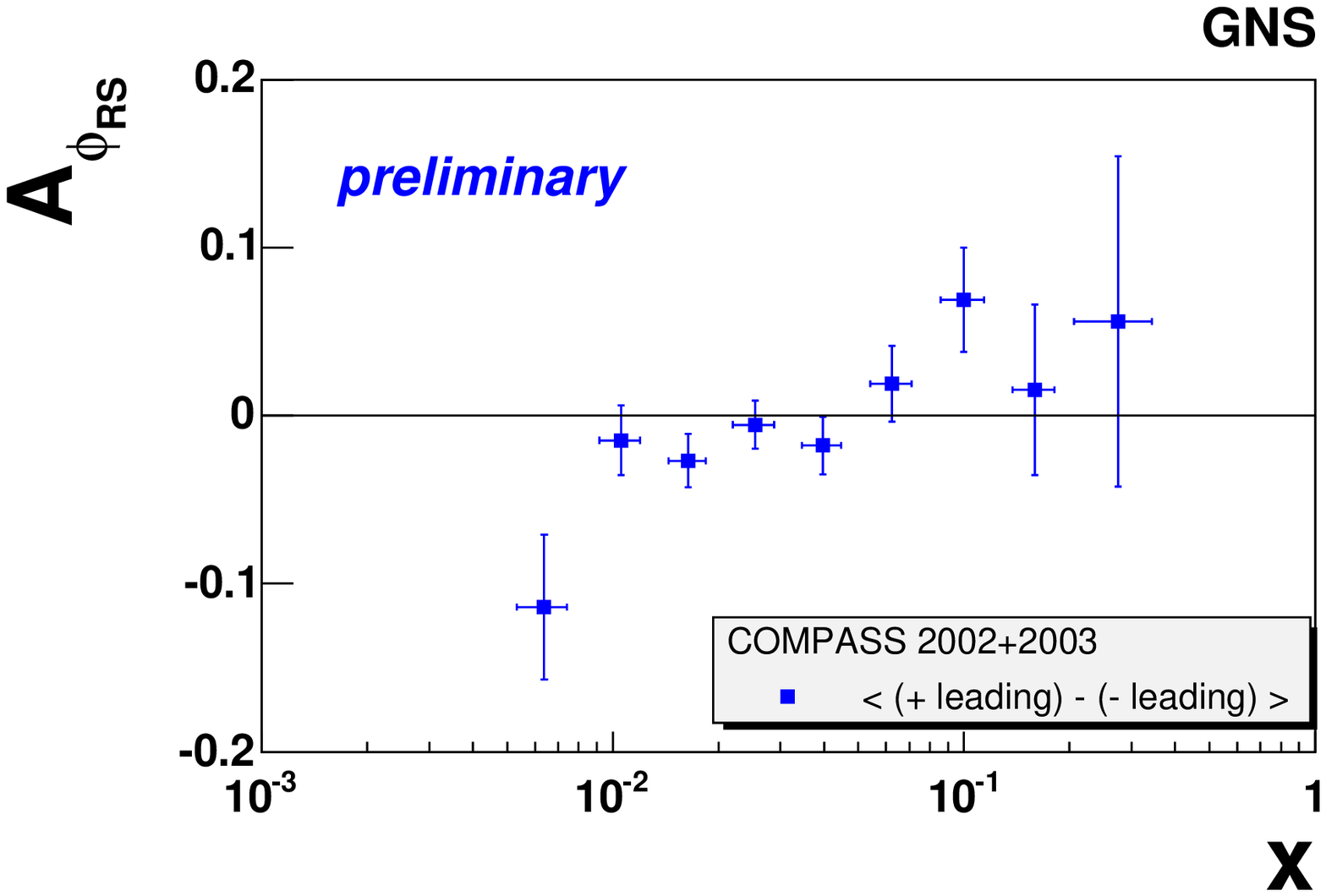}\\
\end{center}
\vspace*{-0.4cm}
{\small{\bf Figure 5.} Two-hadron asymmetry $A_{\phi_{RS}}$
as a function of $x$ for the two-oppositely charged hadrons with 
the largest $p_T$ in the event.
In the upper plot
the asymmetry is shown separately for the case in which the
leading hadron (hadron 1 in the calculation of $\vec{R}_T$) is
positive and  negative.
The lower plot shows the asymmetry $A_{\phi_{RS}}$ for the same
events, but taking as hadron 1 always the positive one.} 
\end{figure}

To search for an effect, we have tried also different selection
of the hadron pairs, in particular we have evaluated the asymmetries
for the two-oppositely charged hadrons with the largest $p_T$ in the
event.
The asymmetry $A_{\phi_{RS}}$ is shown separately for the case of
positive leading hadrons and of negative leading hadrons in 
Fig.~5 (top).
At large $x$ a hint for a mirror symmetry of the two-sets of
data could be seen, suggesting that the possible asymmetry
(if any) could be associated with the hadron charge.
We have weighted averaged the two-sets of data,
 evaluating $\phi_{RS}$ starting always with the positive hadron,
and the resulting asymmetry is shown in Fig.~5 (bottom).
Some effects could be seen at large $x$, but clearly more statistics is 
mandatory before drawing any conclusion.

For all these analyses, particle identification as provided by the RICH 
detector has not been used.
Work is ongoing to evaluate the two-hadron asymmetries for identified
charged hadrons.
The effect of the RICH identification of hadrons is illustrated in Fig.~6,
which refers to the data collected in the year 2003, when the RICH was 
fully operational.
\begin{figure}[h!]
\begin{center}
\begin{tabular}{cc}
\hspace*{-.6cm}
\includegraphics[width=7.cm]{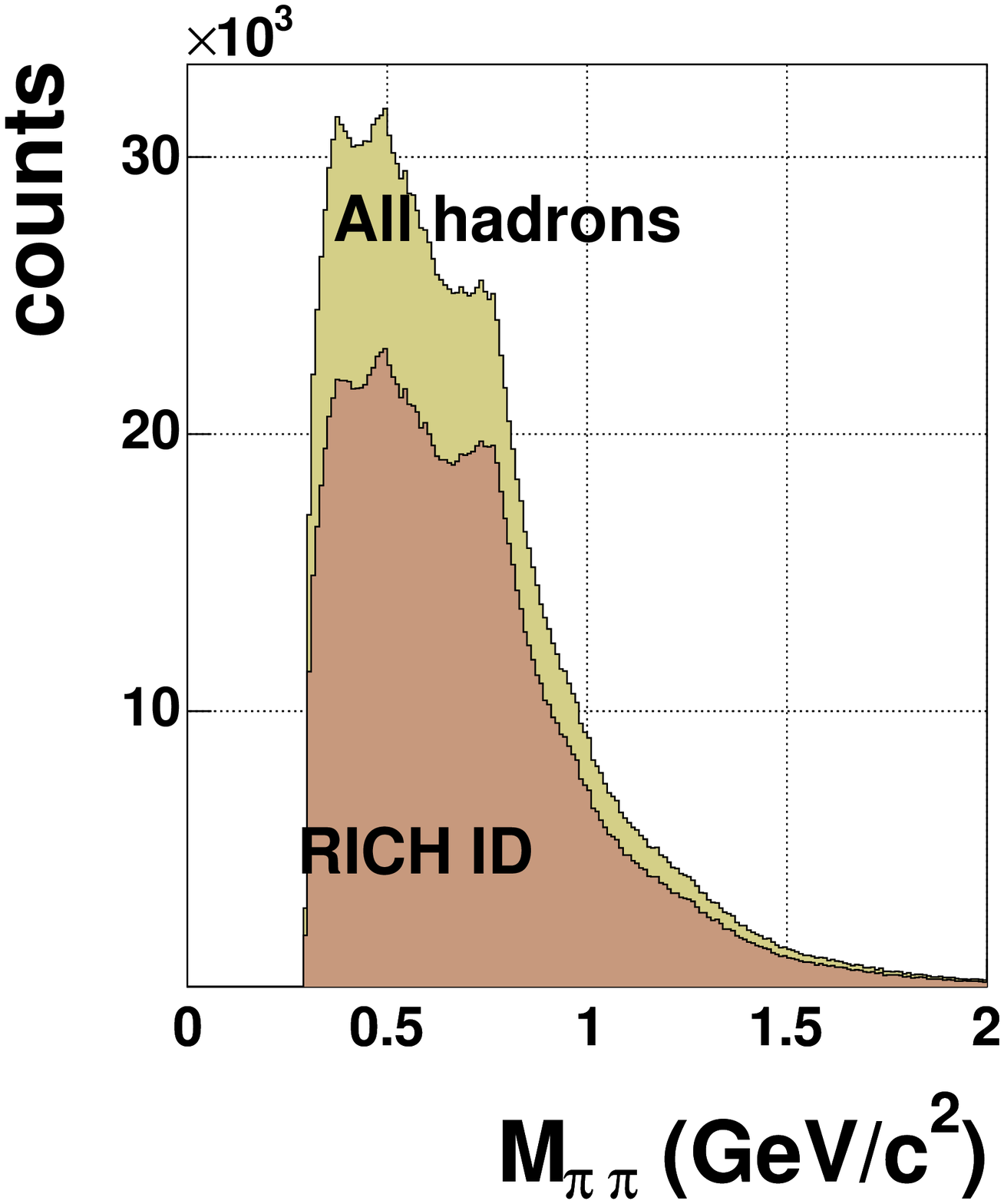}&
\includegraphics[width=7.cm]{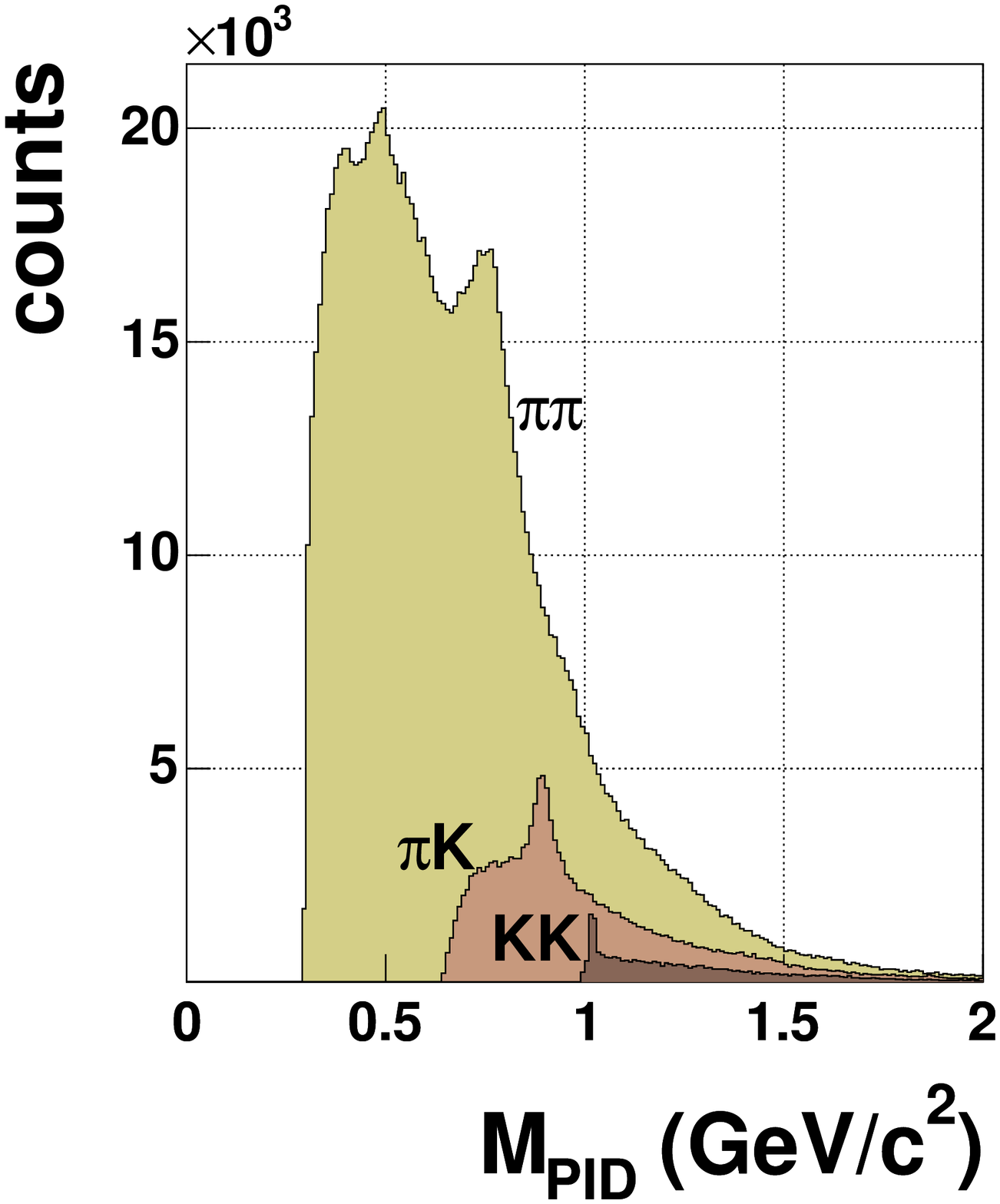}\\
\end{tabular}
\end{center}
\vspace*{-0.4cm}
{\small{\bf Figure 6.} Left plot: invariant mass distribution of two-hadrons 
for all reconstructed hadron pairs and for the hadrons  identified by RICH
(the $\pi$ mass is assumed). 
Right plot: histograms of the invariant masses of hadron pairs with 
the mass hypothesis 
for each hadron as given by the RICH.}
\end{figure}
The left plot compares the
invariant mass distribution  for all reconstructed  
hadron combinations and for the combinations where for both particles 
the RICH has provided an 
identification.
The fraction of the combinations with the RICH identification is $\sim
75\%$ of all 
reconstructed pairs.  
In the plot to the right one can see the effect of hadron identification
in the invariant mass spectrum: 
in the three histograms the 
invariant mass is evaluated following the mass assignments as given by the 
RICH.

\section{$\Lambda$ asymmetries}
Still another approach to transversity is based on the measurement 
of the spin 
transfer to the $\Lambda$ hyperons produced in the DIS
on transversely polarized targets, 
as originally suggested in~\cite{baldracchini,artru,kunne}, 
and more recently by Anselmino~\cite{anselmino}.

If in the fragmentation process at least part of the struck quark 
polarization is transfered to the $\Lambda$, than
the angular distribution in the weak $\Lambda \to p \pi^-$ decay can 
provide information on the initial polarization state of the quark in the 
nucleon.
The $\Lambda$ polarization measured experimentally is therefore given by:
\begin{eqnarray}\label{eq:P_T}
  P_{\Lambda}^{T}\ &=&\ \frac{d \sigma^{\mu N^{\uparrow} \rightarrow 
\mu' \Lambda^{\uparrow} X}\ -\ d \sigma^{\mu N^{\downarrow} \rightarrow 
\mu' \Lambda^{\uparrow} X}}{d \sigma^{\mu N^{\uparrow} \rightarrow 
\mu' \Lambda^{\uparrow} X}\ +\ d \sigma^{\mu N^{\downarrow} \rightarrow 
\mu' \Lambda^{\uparrow} X}}\ =\   
  \nonumber\\ 
&=&\ \displaystyle f P_{T} D_{NN}\cdot
  \frac{\sum_q e_q^2 \Delta_T q(x) \Delta_T D^{\Lambda}_{q}(z)}
{\sum_q e_q^2 q(x) D^{\Lambda}_{q}(z)},
\end{eqnarray}
where the $T$-axis points along the polarization vector of the struck quark, 
and $\Delta_T D^{\Lambda}_{q}(z)$ is the polarized fragmentation function 
(chirally-odd) that 
describes the spin transfer from the quark to the final state hyperon.
The quantities
$f$, $P_{T}$, and $D_{NN}$ are the same as in equation~\ref{eq:colsiv}.

The event selection is based on the requirement to have a scattering 
$\mu \rightarrow \mu'$ primary vertex reconstructed within the geometrical
volume of the target, together with a two-body charged decay of a neutral 
particle downstream of the target.
The $\Lambda$ hyperons undergo the decay $\Lambda \to p \pi^-$ in about 
63\% of the cases. 
The decay is detected as a V-shaped vertex in the reconstructed events.

The main sources of background in the $\Lambda$ sample come from $K^0$ decays, 
photon conversion and fake vertices from accidental track associations. 
To reduce the background, the longitudinal position of the decay vertex is 
restricted to a region between the target exit window and the first tracking 
station. 
A minimal transverse momentum $p_T > 23$~MeV/c of the decay proton with 
respect to the decaying hyperon is required.
The invariant mass spectrum is shown in Fig.~7: only
events with a reconstructed $\Lambda$ invariant mass in a window
of 0.70 GeV/c$^2$ centered on the PDG value of the $\Lambda$ mass, are kept in 
the final event sample used for the polarization calculation. 
This analysis has used all the data collected with the transversely
polarized deuteron target in 2002 and 2003.
The overall number of detected $\Lambda$ decays in this sample is about 20000. 
\begin{figure}[bt] %
\begin{center}
\includegraphics[width=7.8cm]{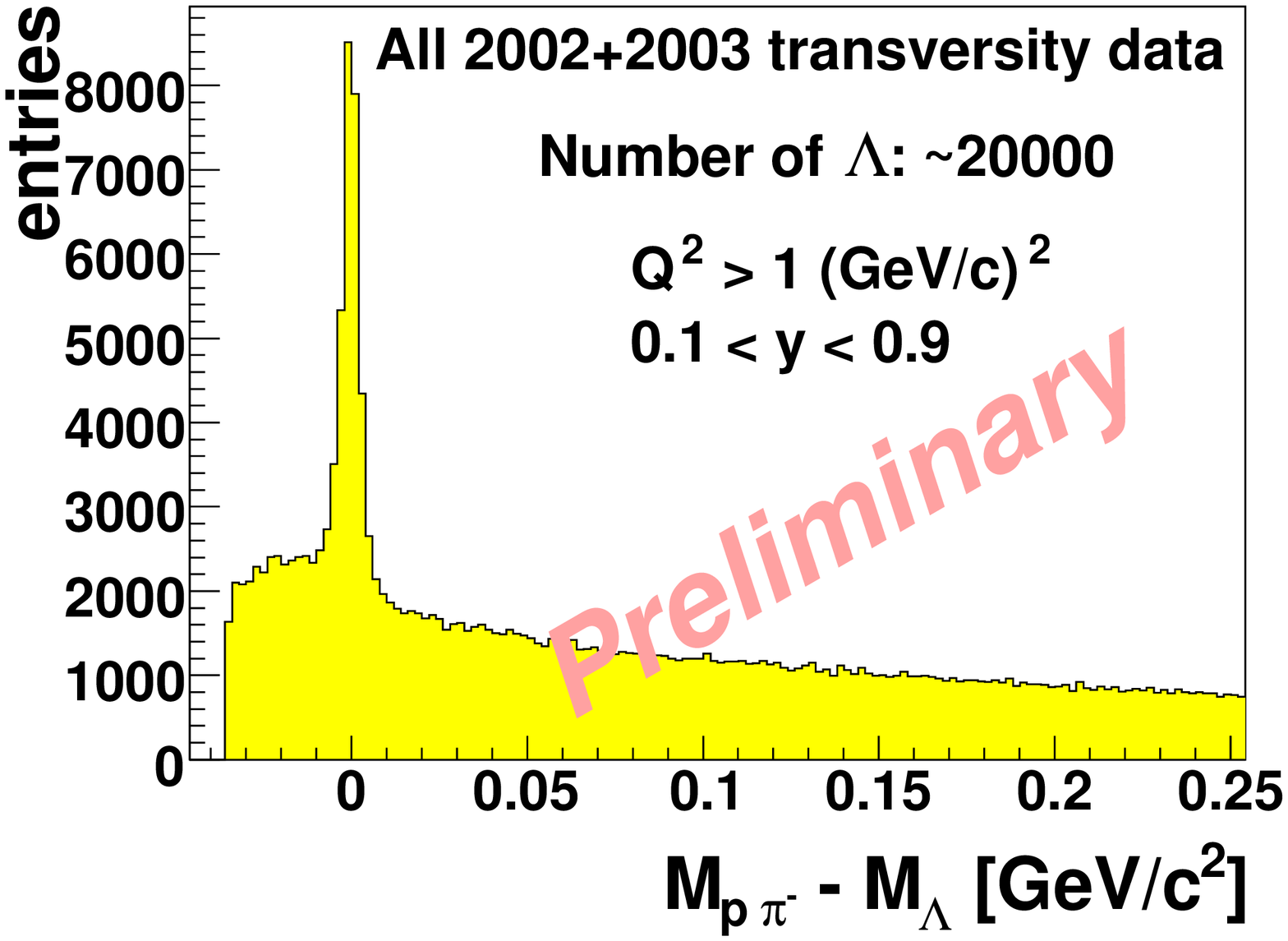}\\
\end{center}
\vspace*{-0.4cm}
{\small{\bf Figure 7.} Invariant mass spectrum of the data 
sample used in this analysis,
 after all the event selection cuts. The overall number of detected $\Lambda$ 
decays is about 20000.}
\end{figure}

The angular distribution of the decay proton in the $\Lambda$ rest frame, 
measured in the experiment, is given by
\begin{eqnarray}
  \frac{dN}{d\theta^{\ast}_T} & = & N_0\cdot\left(1+\alpha P^T_{\Lambda} 
\cos(\theta^{\ast}_T)\right) \cdot Acc(\theta^{\ast}_T)
\end{eqnarray}
where $\theta^{\ast}_T$ is the angle of the proton with respect to
the $T-$axis.
The $Acc(\theta^{\ast}_T)$ function represents the distortion of the 
theoretical angular distribution introduced by the experimental apparatus. 
This distortion is usually corrected by combining real data and MonteCarlo 
(MC) simulations. This approach is however quite sensitive to the accuracy 
of the MC description of the experiment, and requires large MC data samples 
to get a good statistical accuracy.

In this analysis we used a technique based only on real data samples, 
exploiting some of the symmetries of the experimental apparatus.
The technique is based on the combination of two-data taking periods, in 
the same experimental conditions but with opposite target cell orientations. 
Under general assumptions on the existing symmetries, the acceptance functions 
are cancelled and only the terms proportional to the true $\Lambda$ 
polarization remain.
The method is described in detail in Ref.~\cite{aferrero},
and will not be repeated here.

The measured $P_{\Lambda}^T$ as a function of $x$ is shown in 
Fig.~8 (left)  for the 
full data sample and in Fig.~8 (right) for the DIS region 
($Q^2 > 1$~(GeV/c)$^2$). The measured values are compatible with zero in 
all the accessible $x$ range. 
The data points at $x \sim 0.1$, were the 
transversity distribution function 
is expected to be peaked, still have a poor statistical accuracy, therefore 
no conclusion can be drawn yet on the spin transfer from the target 
quark to the final state $\Lambda$.
\begin{figure}[bt]
\begin{center}
\begin{tabular}{cc}
\hspace*{-.7cm}
\includegraphics[width=6.8cm]{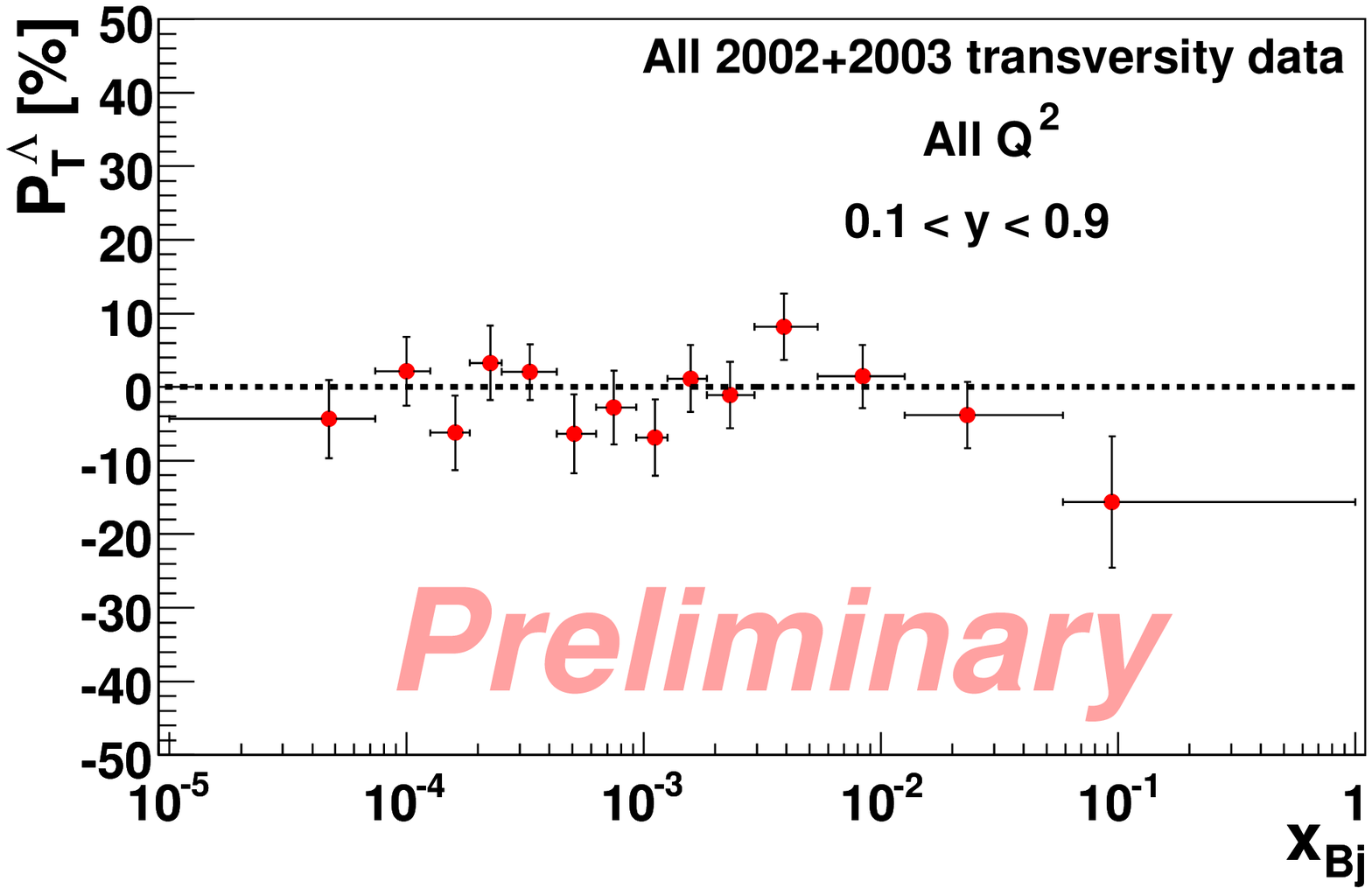} &
\includegraphics[width=6.8cm]{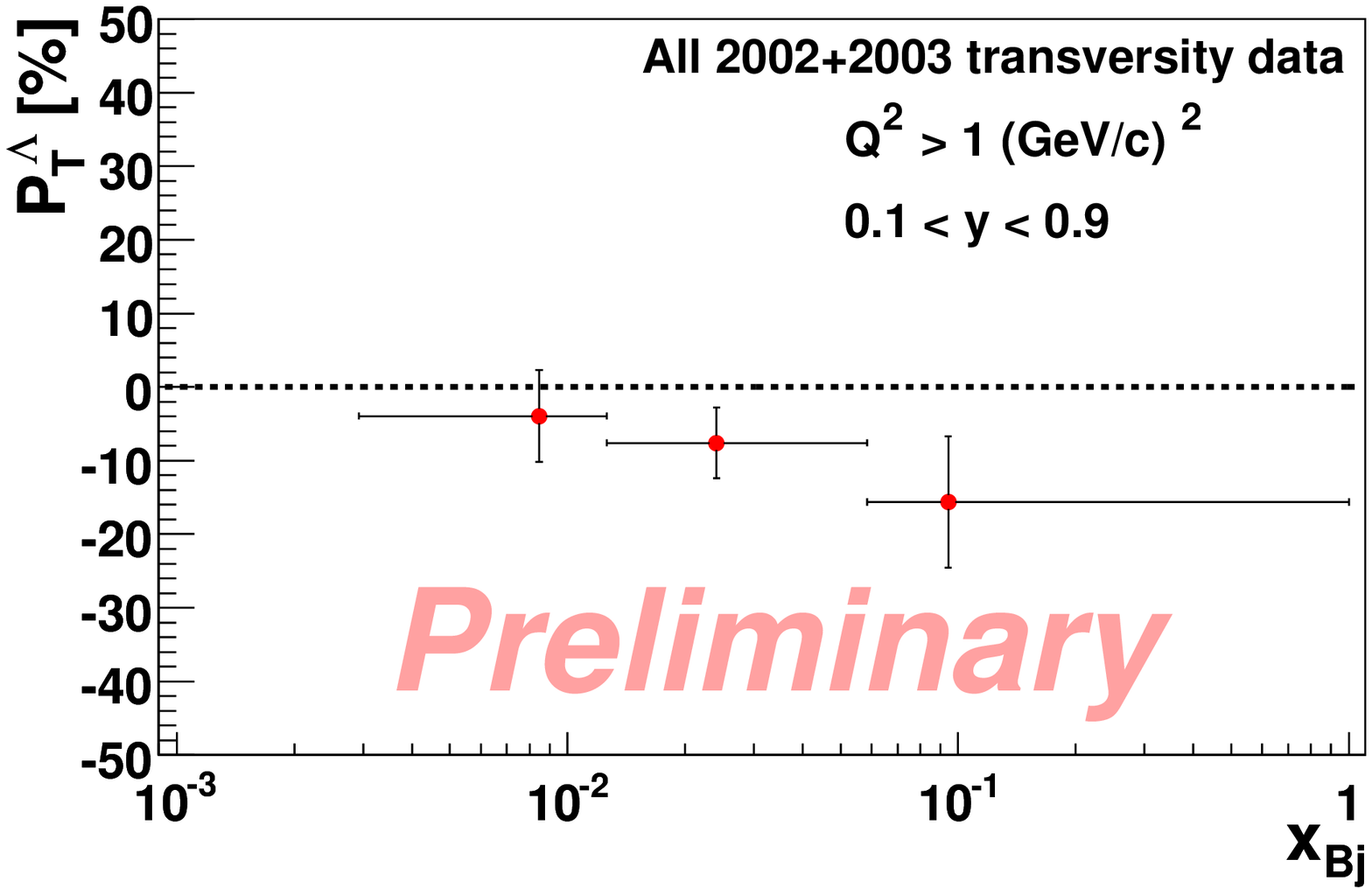}\\
\end{tabular}
\end{center}
\vspace*{-0.4cm}
{\small{\bf Figure 8.} Measured $\Lambda$ polarization as a function of $x$, 
without (left) and with (right) the cut $Q^2 > 1$(GeV/c)$^2$.} 
\end{figure}

\section{Conclusion and outlook}
It has been a pleasure to be again in Dubna and attend this most interesting
workshop, and I'd like to thank Anatoly and the Organizing Committee for
the invitation.

I have tried to summarize the physics motivations we have in COMPASS
to pursue the physics case for transversity.
The case for transversity has been in COMPASS from its very beginning,
and has gained momentum as time has passed.
Due to the variety of physics issues the Collaboration wants to address,
we could dedicate only part of the running time to transverse
polarization, and given the importance of measuring
$\Delta G /G$ data have been taken in so far with the deuteron target,
whose figure of merit is particularly good for that measurement.
Still, we have produced the first ever results of Collins and
Sivers single spin asymmetries on a transversely polarized
deuteron target.

I have reported on transversity data collected over three years,
2002, 2003, and 2004 (this year the experiment is on pause,
all machines at CERN being switched off),
and described the three approaches we have to transversity:
single hadron asymmetries, two-hadron asymmetries, and
$\Lambda$ polarization.
Within statistics, the present evidence from 2002 and 2003
data is that all the measured asymmetries on the deuteron
are compatible with zero.
Taking into account the HERMES results on a proton target
(non-zero effects by many standard deviations) and the BELLE
result on $e+e^- \rightarrow$ hadrons (convincing evidence for
non-zero Collins effect) we can only conclude that there
are cancellation effects between protons and neutrons.

At this point it is of the greatest importance to reduce as much as
possible the errors of our measurements on deuterons,
and we are finalizing the analysis of the collected data.
The next step is to measure transversity with the transversely polarized 
proton target, which is in the plans of our Collaboration for 2006.
This measurement will take advantage of the new COMPASS polarized 
target magnet, which will allow to increase the acceptance
particularly at large $x$, where transversity is expected to be
larger.


%

\end{document}